\documentclass[showpacs,preprintnumbers,amsmath,amssymb]{revtex4}
\usepackage{graphicx}
\usepackage[latin1]{inputenc}
\begin{document}
\preprint{IMAFF-RCA-09-02}
\title{The entangled accelerating universe}

\author{Pedro F. Gonz\'{a}lez-D\'{\i}az and Salvador Robles-P\'{e}rez}
\affiliation{Colina de los Chopos,  Instituto de F\'{\i}sica
Fundamental,\\ Consejo Superior de Investigaciones
Cient\'{\i}ficas, Serrano 121, 28006 Madrid (SPAIN) and Estación
Ecológica de Biocosmología, Pedro de Alvarado, 14, 06411-Medellín,
(SPAIN)}
\date{\today}

\begin{abstract}
Using the known result that the nucleation of baby universes in
correlated pairs is equivalent to spacetime squeezing, we show in
this letter that there exists a T-duality symmetry between
two-dimensional warp drives, which are physically expressible as
localized de Sitter little universes, and two dimensional
Tolman-Hawking and Gidding-Strominger baby universes respectively
correlated in pairs, so that the creation of warp drives is also
equivalent to spacetime squeezing. Perhaps more importantly, it
has been also seen that the nucleation of warp drives entails a
violation of the Bell's inequalities, and hence the phenomena of
quantum entanglement, complementarity and wave function collapse.
These results are generalized to the case of any dynamically
accelerating universe filled with dark or phantom energy whose
creation is also physically equivalent to spacetime squeezing and
to the violation of the Bell's inequalities, so that the universe
we are living in should be governed by essential sharp quantum
theory laws and must be a quantum entangled system.
\end{abstract}

\pacs{04.60.-m, 98.80.-k, 03.67.Bg}

\maketitle

There exist many kinds of spacetime entities which are denoted by
the term universe. The so-called Friedmann-Robertson-Walker
universes, the parallel universes, the De Sitter universe or the
baby universes Wick rotated from Euclidean wormholes, to quote
just a few. In this report we shall show however that all of such
universes correspond actually to a unique scale-invariant quantum
cosmic scenario that has no classical counter-part. It will be
also shown that such a cosmic scenario is equivalent to violating
the Bell's inequalities [1] and, therefore, talking about any of
the conventional models for quantum cosmology is rather redundant
and indeed meaningless because quantum theory and cosmology are
actually equivalent descriptions of the same deep physical
reality.

Several years ago the idea was advanced that the nucleation of
correlated baby universes, taken to be the Lorentzian sector of
Euclidean wormholes, is equivalent to squeezing the spacetime [2].
These baby universes can first be represented as Tolman-Hawking
closed spaces which are described by the metric
\begin{equation}
ds^2=a(\eta)^2 \left(-d\eta^2 +d\Omega_3^2\right) ,
\end{equation}
where $\eta=\int dt/a(t)$ is the conformal time, $d\Omega_3^2$ is
the unit metric on the three-sphere, and $a(\eta)$ is the scale
factor
\begin{equation}
a_b (\eta)=R_0 \cos\eta ,
\end{equation}
with $R_0$ the maximum radius of the baby universe and $a_b
(t)=\sqrt{R_0^2-t^2}$.

Now, it has been more recently shown [3] that the two-dimensional
metric of a warp drive can be expressed in terms of the conformal
time in the manifestly cosmological form
\begin{equation}
a_w (\eta)=\frac{R'_0}{\cos\eta} ,
\end{equation}
where $R'_0$ is the maximum radius of a spatially closed de Sitter
like local space.  We may be therefore uncovering an $a\rightarrow
1/a$ duality symmetry between two-dimensional warp drives and
two-dimensional baby universes which, if confirmed to hold, would
entitle us to accomplish the conclusion that the creation of warp
drives is also equivalent to spacetime squeezing, such as it is
currently believed [4]. That this is actually the case can be
checked by showing that the purely gravitational part of the
Hilbert-Einstein two-dimensional action corresponding to baby
universes and warp drives is the same. Generally, for the relevant
geometric sector of the two-dimensional Hilbert-Einstein action
one may write
\begin{equation}
S=M_p^2\int dx^2 a^2 R-2M_p^2\int dxaTrK ,
\end{equation}
in which $M_p$ is the Planck mass, $R\equiv R(a)$ is the Ricci
curvature scalar and $K$ is the second fundamental form on the
one-dimensional boundary. Computing this action sector for the
scale factors (2) and (3) the result is immediately derived that
such an action is in fact the same for both metrics and given by
\begin{equation}
S_b =S_w =6M_p^2 S_2 \tan\eta ,
\end{equation}
with $S_2$ the proper surface of the given baby (or warp)
universe. It then follows that, relative to the geometric part of
the action, the $a\rightarrow 1/a$ duality symmetry holds for
two-dimensional baby and warp universes having any relative sizes,
and therefore, since most of the physics of these two spacetime
constructs is concentrated on two dimensions, all of their
observable physical properties are undistinguishable from one
another and hence they are both equivalent to squeezing the
spacetime. Now, since squeezing is a quantum phenomenon devoid of
any classical counterparts [5], it follows that the baby universe
spacetime and the warp universe spacetime also are both quantum in
nature, such as it must happen with the scale-factor duality
symmetry between them.

The case for Giddings-Strominger axionic baby universes [6] is a
little more difficult to show but from the very onset we know that
it is at the end of the day transformable in the one for the
Tolman-Hawking case [7] and hence one would expect it to have the
same properties. The metric of a Giddings-Strominger baby universe
is given as in Eq. (1), with
\begin{equation}
a_{GS}=R_0^{GS}\cos^{1/2}(2\eta) .
\end{equation}
There would then exist a solution such that
\begin{equation}
a_d (\eta)=\frac{R_0^d}{\cos^{1/2}(2\eta)}
\end{equation}
whose two-dimensional version must be dual to the two-dimensional
version of metric (6). The corresponding geometric sectors of the
Hilbert-Einstein actions are in fact the same and again given by
expression (5). In terms of the Robertson-Walker time the scale
factor (7) can be expressed as an elliptic function
\begin{equation}
a(t)=R_0^d {\rm nc}\left(\frac{\sqrt{2}t}{R_0^d}\right)
\end{equation}
with $0<t<R_0^d K(1/\sqrt{2})/\sqrt{2}$ and $K(x)$  the complete
elliptic integral of the first kind. However, the metric given by
Eq. (8) in terms of time $t$ does no longer describe a
Friedmann-Robertson-Walker metric as it can be checked by
embedding this two-dimensional spacetime as the three-hyperboloid
\begin{equation}
-T^2 + S^2 + X^2 = R_0^{d2} ,
\end{equation}
with the Lorentzian metric
\begin{equation}
ds^2 = -dT^2 + dS^2 + dX^2 .
\end{equation}
This embedding can be achieved by exhibiting the new coordinates
in terms of the elliptic functions ${\rm sc}$ and ${\rm nc}$ in
the form
\[T=R_0^d {\rm sc}\left(\frac{\sqrt{2}t}{R_0^d}\right)\]
\begin{equation}
S=R_0^d {\rm nc}\left(\frac{\sqrt{2}t}{R_0^d}\right)\sin\rho
\end{equation}
\[X=R_0^d {\rm nc}\left(\frac{\sqrt{2}t}{R_0^d}\right)\cos\rho\]
with which we in fact get a manifestly non
Friedmann-Robertson-Walker metric
\begin{equation}
ds^2 = -2{\rm dc}^2\left(\frac{\sqrt{2}t}{R_0^d}\right)dt
+R_0^{d2}{\rm nc}^2\left(\frac{\sqrt{2}t}{R_0^d}\right)d\rho^2 ,
\end{equation}
${\rm dc}$ being still another elliptic function. A
Friedmann-Robertson-Walker metric can then be obtained by
re-defining the time so that
\begin{equation}
\theta = \sqrt{2}\int{\rm
dc}\left(\frac{\sqrt{2}t}{R_0^d}\right)dt = R_0^d \left[{\rm
nc}\left(\frac{\sqrt{2}t}{R_0^d}\right) + {\rm
sc}\left(\frac{\sqrt{2}t}{R_0^d}\right)\right]
\end{equation}
with which the metric becomes finally
\begin{equation}
ds^2 = -d\theta^2 + R_0^{d2} \cosh^2
\left(\frac{\theta}{R_0^d}\right) ,
\end{equation}
where $0<\theta<\infty$. Metric (14) can then be again interpreted
as that for a two-dimensional warp drive and hence we again derive
the same result as for a Tolman-Hawking baby universe.

Let us now consider the possible connection between baby universes
and warp drives with the essential property behind quantum
entanglement, complementarity and wave function collapse, that is
the Bell's inequalities [1]. Starting with a detailed comparison
of the original intentions of Bohr and Einstein in their
development of quantum mechanics and general relativity, Sachs
showed [8] that the goals of general relativity are more
insightful and subsume those of quantum mechanics, with "general
relativity playing the role of a forest and quantum mechanics that
of its trees". In what follows we shall use the properties
discovered above in order to investigate whether localized warp
drives cropped up in the universe are also connected to a
violation of the Bell's inequalities [1]. If such a task led to
that connection then the very much dismissed Einstein's dream that
it is general relativity where the deepest roots of quantum theory
reside [8] would be re-opened and mark one more example of the
tremendous Einstein insight, leading this time to a new avenue to
unity quantum mechanics and gravitation.

Violation of Bell's inequalities is attained when the following
inequality holds [5]
\begin{equation}
C=\frac{\langle a^{\dagger}aa^{\dagger}a\rangle}{\langle
a^{\dagger}aa^{\dagger}a\rangle+
\langle\left(a^{\dagger}\right)^2\left(a\right)^2\rangle} \geq
0.707 ,
\end{equation}
where the $a$'s are Fock annihilation and creation quantum
operators.

Now, from inequality (15) and the definition
\begin{equation}
g_n^{(2)} =\frac{\langle n^2\rangle -\langle n\rangle}{\langle
n^2\rangle} \geq 1-\frac{1}{\langle n\rangle}
\end{equation}
where use has been made of  the condition $\langle
n^2\rangle/\langle n\rangle \geq 1$, we get
\begin{equation}
C=\frac{1}{1+\frac{\langle n\rangle^2}{\langle
n^2\rangle}g_n^{(2)}}\geq\frac{1}{1+g_n^{(2)}} .
\end{equation}

We compute then the master equation for the second-order
correlation function from the matrix elements in the baby universe
Fock space of the matter field number states [2], in the diagonal
representation
\begin{equation}
\dot{\bar{P}}_n (k,t) =-8\left(n+\frac{1}{2}\right)^2
\left(N+\frac{1}{2}\right)\sinh(2k_0)\bar{P}_n (k,t) ,
\end{equation}
in which $N=0,2,4,...$ denotes the initial number of baby
universes, $\sqrt{2k}$ is the proper distance on the wormhole
inner 3-manifold between the two correlated points at which two
baby universes are created or annihilated, and
$k_0=\sqrt{2k/(R_0^2-k)}$, with $R_0$ is the smallest value of the
scale factor in the connected manifold, to obtain [2]
\begin{equation}
\dot{g}_n^{(2)} = P(N,k_0)\left[\frac{1}{4}\langle n\rangle
g_n^{(2)2}+\frac{1}{4}\left(8\langle n\rangle^2
g_n^{(3)}-7\right)g_n^{(2)}\right.\left. -\langle
n\rangle\left(\langle n\rangle g_n^{(4)}+6g_n^{(3)}\right)\right]
\end{equation}
where $g_n^{(3)}$ and $g_n^{(4)}$ are the third- and fourth-order
coherence functions, respectively, and
$P(N,k_0)=8\left(N+\frac{1}{2}\right)\sinh(2k_0)$ . For the vacuum
case, Eq. (19) admits the exact solution
$g_0^{(2)}(0,k_0)=\exp\left(-\frac{7}{2}P(0,k_0)t\right)$, so that
\begin{equation}
C\geq \frac{1}{1+e^{-\frac{7}{2}P(0,k_0)t}} .
\end{equation}

Thus, for the vacuum case there will always be a large enough time
for which Bell's inequalities are violated. Such a time is smaller
than or as most equal to $t_v =2\ln 2.37/(7P(0,k_0))$. This
conclusion is still valid for small, nonzero values of $\langle
n\rangle$ and even in the limit of large $\langle n\rangle$ where
\[g_n^{(2)} \simeq \frac{1}{2}
\left(1+\exp\left[2P(N,k_0)t\right]\right) .\]

It follows that, though in this limit the right hand side of
expression (20) is in this case defined to be smaller than 0.707
even for $t=0$, the inequality relating it with $C$ can leave
still a residual room for the violation of Bell's inequalities in
the situation where we usually expect the classical limit to hold,
so allowing multiverse descriptions to call for a joint quantum
treatment. The conclusion can then be drawn that warp drives
entail the phenomena of quantum entanglement, complementarity and
wave function collapse. Whether other special kinds of spacetime
involving exotic matter with negative energy would also give rise
to violations of the Bell's inequalities is a matter which
deserves further consideration.

The point now is, provided a de Sitter space by itself implies the
very essential phenomena of quantum entanglement, complementarity
and wave function collapse, would any dynamical generalizations of
a cosmological constant described by a quintessential or
k-essential dark or phantom energy field also entail the deepest
essentials of quantum theory by themselves?.

Let us first assume from the onset that the nucleation of baby
universes in pairs can be equivalently described by means of a
duality symmetry transformation in terms of closed universes
filled with an homogeneous and isotropic fluid, with equation of
state $p=w \rho$, $p$ and $\rho$ being the pressure and the energy
density, respectively, and $w$ a parameter which for the sake of
simplicity we take here to be constant. From the equation of the
cosmic energy conservation,
\begin{equation}
d\rho = - 3 (p + \rho ) \frac{d a}{a} ,
\end{equation}
the solutions to the equation of motion can be computed for such
closed universes. In conformal time $\eta$, they are given by [9]
\begin{equation}
\eta - \eta_0 = \pm \int \frac{d a}{a \sqrt{\lambda_0^2 a^{2 -
3\beta}- 1 } } = \pm \frac{1}{\alpha} \arccos\frac{1}{\lambda_0
a^\alpha} ,
\end{equation}
with $\lambda_0$ a constant, $\beta = 1 + w$ and $\alpha=
1-\frac{3\beta}{2}\neq 0$. Then, the considered baby universes
resulting from the cosmic solutions when the duality symmetry
holds can be taken as those described by the metric (1),
\begin{equation}\nonumber
ds^2 = a^2(\eta) (- d\eta^2 + d\Omega_3^2) ,
\end{equation}
with the scale factor given by
\begin{equation}\label{scale factor3}
 a(\eta) = R_0^{\alpha}\cos^{-\frac{1}{\alpha}}(\alpha\eta) ,
\end{equation}
where $R_0^\alpha \equiv \lambda_0^{\frac{1}{\alpha} }$.

Let us then be concerned with the two dimensional version of this
type of baby universes, for which we take the slice that results
at constant angular variables. As it has been pointed out
previously, it will be now assumed that most of the relevant
physics involved is concentrated on that slice. In such a case,
the duality symmetry $a\rightarrow \frac{1}{a}$ corresponds in Eq.
(23) just to a change of sign in the value of the parameter
$\alpha$. The action given by Eq. (4) becomes thus invariant under
the transformation, $\alpha \rightarrow -\alpha$.

We next notice that there are two special values of the parameter
$w$, which are equivalent to the baby universes which were
considered before. First, the value $w = -1$ (which corresponds to
the case of a positive cosmological constant), i.e., $\alpha = 1$,
is associated by the duality symmetry to the closed Tolman-Hawking
baby universe considered in Eq. (2). The second case, for a value
$w = -\frac{5}{3}$ (which falls well inside the phantom energy
regime [10]), i.e., $\alpha = 2$, corresponds to the
Giddings-Strominger baby universe, given by Eq. (6). There is
still another special value which the parameter $w$ may take on,
$w=-\frac{2}{3}$ (which describes an accelerating universe
dominated by dark energy [11]), i.e $\alpha=1/2$, amounting to a
third kind of Euclidean wormholes characterized by a scale factor
which in its Lorentzian sector is given by
\begin{equation}
a(\eta)=M\cos^2(\eta/2).
\end{equation}
An Euclidean wormhole solution of the form
$a\propto\cosh(\eta_E/2)$ can be derived from the Euclidean
Friedmann-Robertson-Walker Einstein equations by simply adding an
extra quantum term arising from the insertion of a minimum
resolution distance in the background theory, in the case that no
cosmological constant be included [12]. Now, similarly to how we
have shown that the two two-dimensional cosmological solutions
respectively associated with Tolman-Hawking and
Giddings-Strominger two-dimensional baby universes are both
convertible into the two-dimensional warp drive space-time, one
would expect the two-dimensional version of the cosmic scale
factor associated with a two-dimensional baby universe given by
Eq. (24) to be convertible into that for the two-dimensional warp
drive, too. This expectation arises from the result [13] that the
three kinds of baby universe considered above correspond to the
only three existing Euclidean wormhole solutions and it was shown
that they are physically equivalent to each other. In fact, the
two-dimensional cosmic solution for the scale factor
$a(\eta)=\frac{M}{\cos^2(\eta/2)}$ is once again physically
equivalent to that of a two-dimensional warp drive as it can be
readily shown that such a metric is conformal to the one
associated with a closed de Sitter space when the former metric is
expressed in terms of a Friedmann-Robertson-Walker time. Thus,
defining first the time $t=\int\frac{Md\eta}{\cos^2(\eta/2)}$ and
hence, $a(t)=M+t^2/(4M)$, and then a new time $\theta$ through
$t=2M\sinh(\theta/4M^2)$, we have the well-defined two-dimensional
conformally Friedmann-Robertson-Walker line element
\begin{equation}
ds^2=\cosh^2(\theta/4M^2)\left[-d\theta^2 + M^2
\cosh^2(\theta/4M^2)\right],
\end{equation}
with $0\leq\theta\leq\infty$.

When combined with the above discussion on the violation of Bell's
inequalities, if we take into account the restriction that $\alpha
>0$ which is required in order to obtain baby universe solutions,
then the previous results leads inexorably to the conclusion that
always we have a universe which expands in an accelerated fashion,
no matter whether it is phantom, de Sitter or dark energy
dominated, it entails the deepest essentials of quantum theory,
meaning that such a universe, by itself, is a quantum, entangled
system which has no classical analog whatsoever. The rather
bizarre implication that the universe where we live in is of
necessity a quantum universe appears to be made less surprising,
at least when phantom energy is considered. In fact, the very
essential features of a universe filled with such a kind of vacuum
energy mark rather quantum footprints, which manifests in the fact
that the parameter of its equation of state must be quantized and
that the phantom energy density increases with time to tend to a
classical singularity (which likely be smoothed out by quantum
effects) in a finite time in the future. The feature that all
accelerating ways to expand are at the end of the day physically
equivalent makes then the above conclusion quite less bizarre.

Indeed, if the ultimate cause for the current speeding-up of the
universe is a universal quantum entanglement, then one would
expect that the very existence of the universe implied the
violation of the Bell's inequalities and hence the collapse of the
superposed cosmic quantum state into the universe we are able to
observe, or its associated complementarity between cosmological
and microscopic laws, and any of all other aspects that
characterize a quantum system as well. The current dominance of
the resulting quantum repulsion over attractive gravity started at
a given coincidence time would then mark the onset of a new {\it
quantum} region along the cosmic evolution, other than that
prevailed at the big bang and early primeval universe, this time
referring to the quite macroscopic, large universe which we live
in. Thus, quite the contrary to what is usually believed, quantum
physics not just govern the microscopic aspect of nature but also
the most macroscopic description of it in such a way that we can
say that current live is forming part of a true quantum system.

\acknowledgments The authors are very much grateful to Carmen L.
Sigüenza for useful conversations and the Estación Ecológica de
Biocosmología of Medellín, Spain, where most of this research was
carried out. This work was supported by MEC under research project
no. FIS2008-06332/FIS.

\end{document}